# AI and the Future of Digital Public Squares

Beth Goldberg[1,2], Diana Acosta-Navas[3], Michiel Bakker[4,5], Ian Beacock[1], Matt Botvinick[4], Prateek Buch[6], Renée DiResta[7], Nandika Donthi[8], Nathanael Fast[9], Ravi Iyer[9], Zaria Jalan[5], Andrew Konya[10], Grace Kwak Danciu[11], Hélène Landemore[2], Alice Marwick[12], Carl Miller[13], Aviv Ovadya[14], Emily Saltz[1], Lisa Schirch[15], Dalit Shalom[16], Divya Siddarth[17], Felix Sieker[18], Christopher Small[1], Jonathan Stray[19], Audrey Tang[17], Michael Henry Tessler[4], Amy Zhang[20]

[1] Jigsaw, Google
[2] Yale University
[3] Loyola University Chicago
[4] Google DeepMind
[5] Massachusetts Institute of Technology
[6] UK Policy Lab
[7] Georgetown University
[8] Reddit
[9] University of Southern California
[10] Remesh
[11] Independent
[12] Data & Society
[13] Demos
[14] AI & Democracy Foundation
[15] University of Notre Dame
[16] The New York Times
[17] Collective Intelligence Project
[18] Bertelsmann Stiftung
[19] University of California Berkeley
[20] University of Washington

## Abstract

Two substantial technological advances have reshaped the public square in recent decades: first with the advent of the internet and second with the recent introduction of large language models (LLMs). LLMs offer opportunities for a paradigm shift towards more decentralized, participatory online spaces that can be used to facilitate deliberative dialogues at scale, but also create risks of exacerbating societal schisms. Here, we explore four applications of LLMs to improve digital public squares: collective dialogue systems, bridging systems, community moderation, and proof-of-humanity systems. Building on the input from over 70 civil society experts and technologists, we argue that LLMs both afford promising opportunities to shift the paradigm for conversations at scale and pose distinct risks for digital public squares. We lay out an agenda for future research and investments in AI that will strengthen digital public squares and safeguard against potential misuses of AI.

## 1. Introduction

These co-authors, alongside 50 other leading thinkers and technologists, convened on April 3, 2024 in New York City to discuss the potential for recent innovations in AI capabilities to transform the public square and democratic process. This paper advances key insights from that convening alongside applied research on improving digital public squares globally.  The convening and this paper are structured around four areas in which AI-enabled technologies have demonstrated potential to significantly disrupt—and potentially improve—conversations at scale:

1. Collective dialogue systems
2. Bridging systems
3. Community moderation
4. Proof-of-humanity systems

## 1.1 Origins of the Digital Public Square

A "public square" traditionally refers to an open, communal space where people gather for various social, political, and cultural activities. Public squares are a focal point globally for community interaction and expression, facilitating everything from casual socializing to organized events where people can exchange both ideas and goods. The term "public square" has expanded beyond physical gathering points to include the whole set of spaces—physical and online—that fulfill these roles in modern society.

Jürgen Habermas wrote about the "public sphere" as a two-track system comprising both a *formal* and an *informal* track. The metaphorical "public square" is really the informal track of the public sphere, the place where "deliberation in the wild"—unconstrained communication between groups and individuals—takes place in a way that ends up shaping, constraining, and ideally setting the agenda for the more formal track of courts, parliaments, senates, and so on, that govern functioning democracies (Landemore 2022b).

The concept of the internet as a public square traces its origins to the early visions of digital communication theorists who compared the burgeoning network to traditional town squares. The 1994 book *The Virtual Community* by Howard Rheingold envisioned the internet as a new frontier for public dialogue and community formation (Rheingold 1994).  Legal scholars like David R. Johnson and David Post, in their seminal article *Law and Borders—The Rise of Law in Cyberspace*, explored the potential of the internet to transcend traditional governmental boundaries, highlighting its capacity to foster a more participatory democracy by enabling global conversations and decentralized governance (D. Johnson and Post 1996). While the metaphor continues to evolve, the throughline is consistent; the internet serves an unprecedented role facilitating public debate, civic engagement, and social movements (Bohman 2000).

By the 2010s, internet studies scholars like Ethan Zuckerman illuminated how the internet was democratizing both the production and sharing of information, enabling global participation in public discourse and expanded access to more diverse voices (Zuckerman 2013). Danielle Allen and Archon Fung expanded on this, noting that the internet's unique communication forms could "enhance inclusion, political equality, public deliberation, and civic engagement" due to its capacity for many-to-many interactions, structured and open-ended communication, and lower barriers to the production and circulation of information (Fung and Shkabatur 2015); (Kahne, Middaugh, and Allen 2015). The design of the internet of the 2010s laid foundations for the kinds of democratic processes that function primarily outside formal political mechanisms.

## 1.2 Defining an Ideal Digital Public Square

By bolstering these informal mechanisms for democratic participation, the internet brought about some of the conditions of what Hélène Landemore (Landemore 2022b) defines as the ideal of an "Open

Democracy." It created open-ended channels for debate, such as Twitter, that improved access to agenda-setting power for ordinary citizens. In a systematic review of the evidence, greater digital media use was correlated with greater political knowledge, expression, and participation internationally – but also greater polarization (Lorenz-Spreen et al. 2023).

By this light, digital public squares have developed many of the ideal qualities of a democratic space that DuBois envisioned in the 1920s, where each individual can contribute uniquely to human knowledge and wisdom (Du Bois 2017). These online spaces also move us closer to Rawls's ideal of public reason, where citizens can justify their political decisions with reasons that fellow citizens, as free and equal individuals, can reasonably accept or reject within a democratic framework (Rawls 2005). The ideal of the public sphere as described by Jurgen Habermas, however, one of free deliberation and political influence through communicative action aimed at mutual understanding and consensus, remains to be fulfilled at scale in online spaces (Habermas 1991).

A wide range of scholars have studied the role of technology in supporting a more ideal digital public square. In their book *Plurality: The Future of Collaborative Technology and Democracy*, Glen Weyl and Audrey Tang lay out the opportunity for digital platforms to enable real-time, widespread participation in governance, allowing citizens to engage directly in decision-making processes (Tang and Weyl 2024). By integrating advanced digital tools powered by AI into the democratic framework, Tang and Weyl propose a model whereby technology acts as a catalyst for creating more responsive, inclusive, and accountable democratic institutions. Digital platforms designed for the health of the public square can and have enabled large groups of people to listen to and better understand both each other and experts. By tapping into the collective "wisdom of the crowd," these technologies serve in Tang and Weyl's model as enabling factors for empowering the digital public square as a space for meaningful and impactful participation (Tang and Weyl 2024).

Researchers at New_Public developed a set of "Civic Signals" that characterize a flourishing digital public space ("Civic Signals" 2023). These four components include:
- **Welcoming** requires a space to be accessible, to feel safe, attractive, and navigable.
- **Connecting** requires a space to be conducive to friendly, cooperative interaction between diverse people.
- **Understanding** requires a space to provide useful information about how the public square works.
- **Engaging in meaningful action**, a key feature of a robust public square, requires a space to provide the public with incentives to participate.

### 1.3 The Unrealized Potential of the Digital Public Square

While the digital public square is an invaluable expansion of offline fora for discourse, there are many ways in which our online sphere has fallen short of the ideals above. At least three categories of the potential of today's digital public squares, undermining democratic participation in the process:

- **Access and Inclusion**: The promise of a universally accessible public square remains more an ideal than a reality. Rural communities, the poor, minority language speakers and other marginalized groups do not have the same access to digital technologies. Additionally, online spaces often become venues for harassment, disproportionately affecting women and members of historically marginalized communities. This has led members of these groups to refrain at times from participating in public discourse (Matias, Simko, and Reddan 2020).

- **Privacy and Surveillance:** Both state and non-state political actors use digital platforms to track and influence citizen behavior on an unprecedented scale, potentially using this data for political manipulation or suppression of dissent. Mass surveillance and privacy breaches undermine democratic freedoms, drive self-censorship, and subvert the function of digital public squares (Kaye 2019).
- **Distorted Information Ecosystems:** Information is predominately accessed through privately owned channels and platforms that have competing incentives to present the information, with effects that can distort perceptions of reality:
  - **Engagement Optimization:** Some digital algorithms maximize engagement and promote sensationalist and polarizing content. This can obscure facts and heighten emotions, driving increased societal division and polarization and lead users to perceive public opinion as more polarized than it is in reality (Bail 2022). When systems are optimized for engagement, they can become dominated by small groups of extreme users who are willing to engage more, crowding out the vast majority of more moderate participants (Gillespie 2018; Cunningham et al. 2024).
  - **Echo Chambers**: People sort into homophilous groups and echo chambers on and offline, which can reinforce existing beliefs and limit exposure to the diverse viewpoints vital for a healthy democracy. Digital public squares that become echo chambers can amplify ideological segregation, complicating the public's capacity to engage in free and fair discourse (Guess et al. 2018).
  - **Manipulative Propaganda**: Political actors and other vested interests use digital platforms as arenas for targeted political advertising and manipulation. Modern information technology makes it easier than ever before to identify and exploit wedge issues, distorting public opinion. This manipulation can manifest differently in authoritarian regimes and open societies but is deeply concerning in both (Roozenbeek and van der Linden 2024).
  - **Extremist Movements:** Extremists have long leveraged communications technologies to amplify violent ideologies. Digital public squares can become convenient for these actors to radicalize, recruit, and mobilize individuals, which undermines social cohesion, safety and democracy.

These distortions tend to paint public opinion as more polarized than it really is, amplifying extremes (Lees and Cikara 2021), and degrading trust between groups and between the public and civic institutions. These perceptions further contribute to making public spaces less welcoming and inclusive, and limit our ability to participate meaningfully in civic life (Settle 2018). Together with the rest of the access and inclusion issues, and the threats to privacy from surveillance, the digital public square can be seen as falling short of early techno-enthusiasts' optimistic ambitions.

**Position on the Future of Digital Public Squares**

This is an opportune moment to invest in improving the digital public square, reorienting it toward and hopefully moving it closer to the ideal. Alongside increased accessibility and adoption of digital technologies, advances in AI with large language models (LLMs) have opened up new possibilities for creating deliberative spaces and healthier public squares at scale. This paper will explore applications of new AI-enabled technologies for functions such as facilitating sensitive dialogues, moderating toxic speech, rewarding speech that bridges divides, and synthesizing outcomes of large-scale deliberations.

Each of these families of technology will be discussed in turn, including opportunities afforded by the application of LLMs, risks, and recommendations for civil society, policymakers, researchers and technologists. These sections address the following questions:

1. **Deliberation & Collective Dialogue Systems**: *How can we create more informative, inclusive, and scalable feedback loops between the public and decision makers?*
2. **Bridging Systems**: *How can we architect our sociotechnical systems to reward and elevate the ideas that bring us together?*
3. **Community-driven Moderation**: *How can we empower community leaders to better guide the norms of discourse to support healthy, inclusive online communities?*
4. **Proof of Humanity**: *With advances in AI making it increasingly difficult to tell humans from machines online, how do we balance the competing interests of privacy, free speech, and authenticity?*

## 1. Deliberation & Collective Dialogue Systems

Collective Dialogue Systems[1] (CDS) have recently emerged as tools which sit somewhere between traditional surveys and focus groups, allowing for more of the rich and nuanced feedback of the latter, while being more inclusive, scalable, and amenable to quantitative analysis, like the former (Konya et al. 2023; Ovadya 2023).

Such systems scalably allow a) participants to express themselves in their own words, b) participants to engage in parallel such that everyone gets a "fair hearing", c) iterative interaction by participants (participants can contribute after considering the contributions of others), and d) offer some mechanism for synthesizing and understanding the perspectives of the participant body (Ovadya 2023). Examples of such CDS include Polis, Remesh, Make.org, All Our Ideas, CrowdSmart and CitizenLab.

---

[1] This is one of a number of terms which describe related technologies, including Collective Response Systems, Collective Intelligence Technology, Deliberative Technology, etc. Collective Dialogue Systems as a focus was a deliberate choice by the authors as we believe it captures a specific subtype of these technologies with particular potential for impact and improvement.

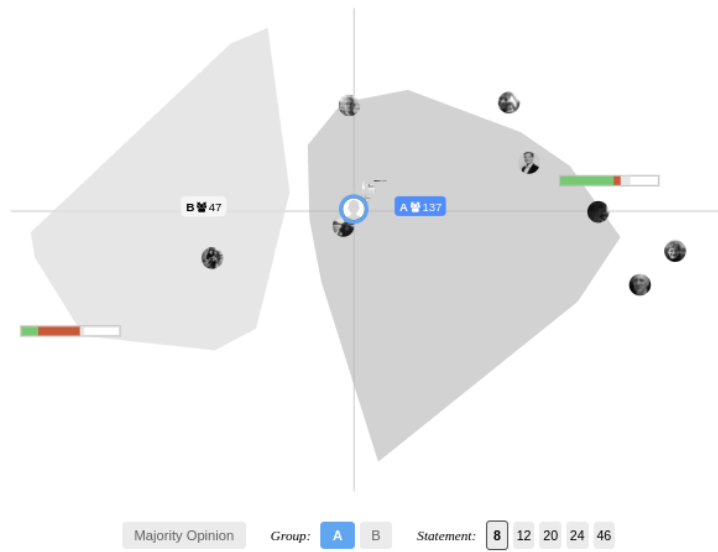

Figure 1: The participation interface for a demonstrative Polis conversation. The figure shows a conversation prompt, a participant-submitted comment presented for vote, an input for submitting a new comment, and a visualization of the emergent opinion landscape including a comment representative of one of the opinion groups.

CDSes have been integrated into formal deliberative decision making processes (New Jersey Office of Innovation 2024), allowing the broader public to directly weigh in on the issues at hand. Participants in CDS-enabled deliberations directly shape the dialogue by both voting on and contributing new ideas, providing decision-makers (and each other) with a better understanding of the full range of perspectives on an issue, expressed in the public's own terms. Frequently, they are able to surface unexpected points of common ground, such as via bridging based ranking algorithms discussed in the following section. This can lead to novel solutions to highly politicized problems, such as climate change policy (Paice and Rausch 2021), medical advice from AI assistants (Konya et al. 2023), and AI governance (S. Huang et al. 2024). CDSes can also be used outside formal decision making to directly reflect back to the public the emergent opinion landscape on a topic and help frame an agenda and boost political engagement (Fishkin et al. 2024). They can thus play an important role in framing and contextualizing political issues ("Bowling Green 2018 Case Study" 2018).

There is evidence that participatory exercises tied to formal decision making processes - from participatory budgeting (C. Johnson, Carlson, and Reynolds 2021) to deliberative polling (Fishkin et al. 2024; "Collective Intelligence Case Studies," n.d.) - lead to increased democratic engagement and satisfaction. This has the effect of bolstering trust in political institutions and reducing misperceptions, protecting the public against the kinds of information attacks and vitriolic content which have contributed to political polarization (Himmelroos and Rapeli 2020).

In formal decision-making contexts, CDSes are typically not stand-alone exercises, but instead operate as part of a larger process that frequently includes in-person deliberations among stakeholders and decision-makers . Sometimes these technologies are coupled with citizen assemblies, where a "jury" of participants is selected to deliberate in person. In these cases, a CDS is often used jointly with citizen assemblies to crowdsource perspectives from the broader public (Paice and Rausch 2021).

These technologies facilitate broad, iterative dialogues that allow the public to deepen their understanding through continuous discussion. Unlike conventional polls and surveys, they provide a platform for participants to express their views, absorb the experiences and ideas of others, identify commonalities, and vote on various proposals (Heinzl et al. 2023). While they typically lack some of the humanizing benefits and depth of reasoned discourse of traditional face-to-face deliberations, they benefit from being less costly to scale out to a larger number of participants, and require less time commitment from participants. They consequently serve a valuable role in creating more opportunities for public engagement in shaping an agenda for the formal track of the public sphere, complementing the adoption of citizen assemblies (Matasick 2020).

## 1.1 Current Applications and Challenges of Collective Dialogue Systems

Collective Dialogue Systems facilitate collective feedback gathering at a scale previously unimaginable. However, they still currently require a significant amount of human labor at every stage of the process. Just understanding enough about these systems to ethically and effectively set up, facilitate and moderate CDS conversations requires a significant amount of training. Gleaning insights at the end of the process is similarly laborious, given how mapping high-dimensional opinion spaces often requires data science skills to distill outcomes for decision makers and participants themselves (C. Small et al. 2021).

The following obstacles presently limit the adoption of CDS and deliberative technologies:

- **Deciphering meaning from data:** While the sensemaking features built into these platforms can often surface valuable insights somewhat quickly, more in depth understanding of the full dimensionality of the emergent opinion landscape typically requires dozens of hours of labor per dialogue, a technical understanding and comfort with statistics, and the ability to write up the results clearly.
- **Education and understanding:** Many people want to run collective deliberations, or are not opposed to doing so, but don't know where to start, what platforms to use, how to set them up, or how to inform and recruit participants. Current onboarding processes are time-intensive and have limited replicability.
- **Cost:** Many of these tools are expensive to host and run. The primary cost of using these tools is rarely the technology itself, but rather the human resources - time and expertise - needed to consult with stakeholders, recruit and inform participants, conduct a CDS, interpret the results, and finally advocate to ensure that decision makers use the insights from the deliberation. Just hosting open source technology can also be prohibitively expensive for some organizations due to the costs of maintaining technical infrastructure, but is often a requirement due to data sovereignty regulations or concerns, which can prevent usage of overseas instances.
- **Institutional capacity and will:** Such a complex, multi-stakeholder undertaking requires significant sustained institutional capacity in terms of staff, time, and political will. A number of factors can imperil an institution's capacity or will to execute a CDS and other participatory exercises, including but not limited to changes in political administration, staff turnover, low or declining public participation in the initiative, lack of political will or capacity to use the dialogue to inform action, or the perception that the CDS was always intended as simply an "experiment" or standalone event. Given the high cost of CDSes - in terms of institutional capacity, political will, as well as financial cost - CDSes are too often one-off experiments (Ryan, Gambrell, and Noveck 2020).
- **Content Moderation**: CDS must be safe spaces for the public to express themselves authentically, necessitating moderation of hateful or toxic rhetoric, which can discourage participation by some. For especially large conversations, it can also become necessary to remove statements that are too similar to each other, to avoid wasting participants' time, and prevent problems with vote sparsity. These tasks require significant human labor, and training to ensure high standards and prevent the censorship of valid topics for debate. Applying LLMs to reduce these human burdens has been discussed by platforms and researchers (T. Huang 2024; Willner and Chakrabarti 2024); the section below on Community-Driven Moderation explores the opportunities and risks of AI-enhanced moderation for CDS and digital public squares more broadly.

## 1.2 Opportunities for AI-enhanced Collective Dialogue Systems

The following opportunities explore how LLMs can address many of the obstacles described above for adopting and scaling CDS. These instructive visions describe how AI can augment peer-to-peer and collective level dialogues, which are gaps in the field of digital participation platforms as identified by Rossello et al (Rossello et al. 2024).

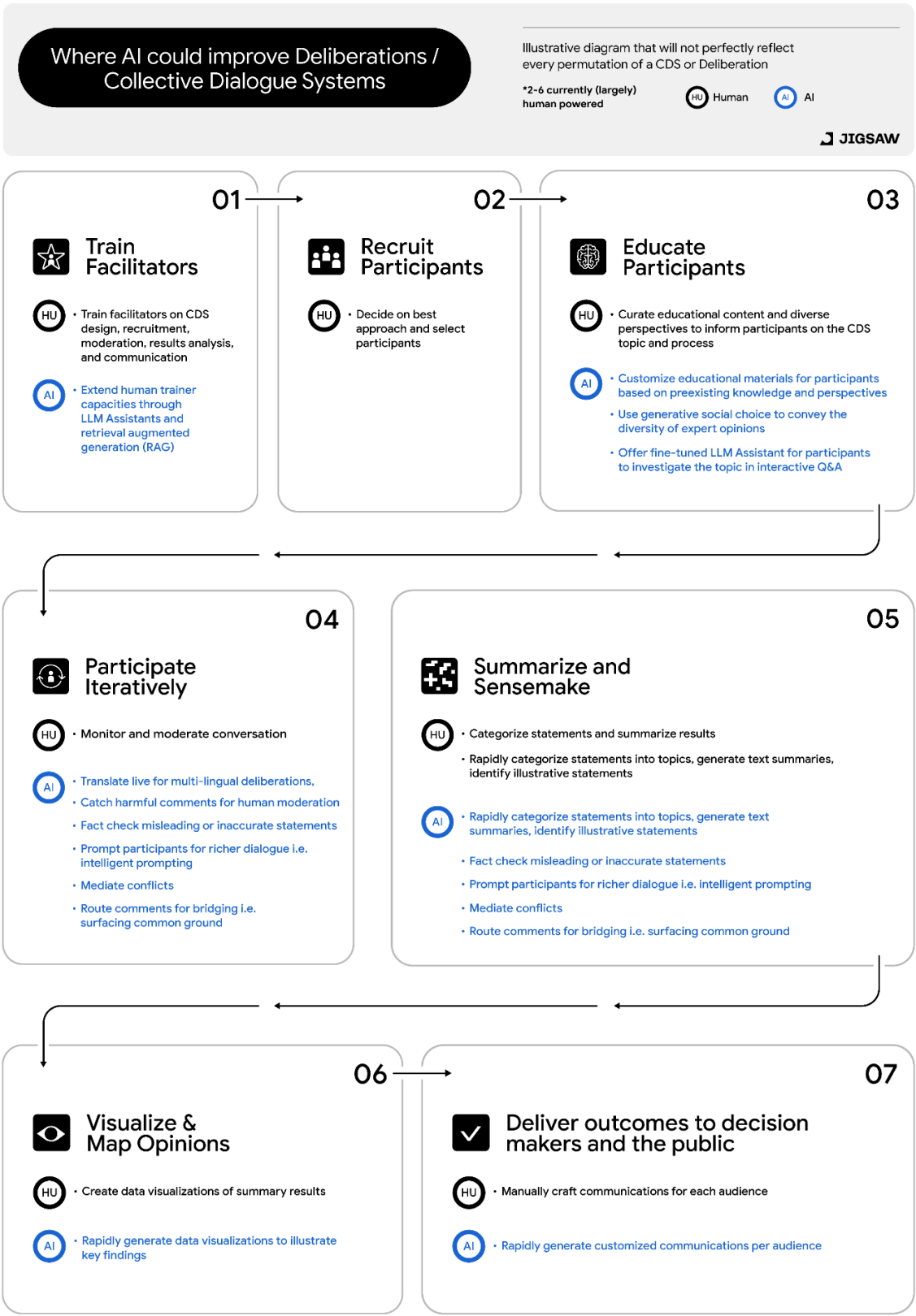

Figure 2: Illustrative overview of a collective dialogue system. While CDSes may extend before and after the seven steps documented here, these steps offer discrete opportunities to integrate AI to enhance CDSes.

### Training Facilitators

Facilitation of both virtual CDS engagements and more traditional in-person deliberations and citizens' assemblies is a highly skilled task that requires significant training. While AI can be used to reduce the burden of many of these tasks, it will likely be quite some time before these tasks may be automated, especially at the interface to systems of formal power and decision making. There is immense value in designing systems that keep humans in the loop at all stages of CDS to ensure that AI behavior is aligned with public interest.

Currently, training individuals in the skills required to set up a CDS, recruit, moderate, and analyze results is a task that falls upon a relatively small community of experts, and takes considerable time and cost. Recent work has looked at how Retrieval Augmented Generation (RAG) systems can be used to augment human intelligence in training tasks (Modran et al. 2024). Applied to this context, an expert-curated RAG system could turbo-charge the training process, and empower skilled facilitators with powerful learning technology to support the next generation of practitioners, especially in historically marginalized populations that may have difficulty funding these training exercises.

### Educational Materials for Participants

A key feature of many well designed deliberative processes is an on-going educational component for the participants, so that they can begin deliberations with a richer understanding of an issue and continue enriching it as the deliberation unfolds. The sourcing of educational material and presentations is often a carefully curated process by experts. This production process itself could be an opportunity to apply LLMs and deliberative methods.

AI can assist with conducting research to understand the topics that participants will be asked to address during deliberations. The GovLab and Citizens Foundation have developed a toolkit called Policy Synth that uses AI agents and genetic algorithms to rapidly conduct large-scale, automated web research on a given topic. The findings from this automated research are then used as the basis for written briefs that frame the discussion topics for participants in a deliberation (Bjarnason, Gambrell, and Lanthier-Welch 2024). This process applies AI to gather, synthesize, and then present complicated evidence into more readily-digestible formats for participants.

Once generated, educational content could be placed in a database where it can be accessed by LLMs (known as *retrieval-augmented generation* or RAG systems), which would allow participants to ask questions about the domain, with the system returning responses citing the source material put together by the experts (Liu et al. 2024). This presents another opportunity for more democratic information curation; generative social choice could be used to balance information from experts with conflicting perspectives by mapping out where experts agree and where there are ambiguities. This contributes to solving the longstanding challenge of better incorporating expert opinion into CDSes and deliberative processes more broadly (Bricker 2024).

### Intelligent Prompting & Mediation

Skilled facilitation involves careful framing to initiate discussions, active listening to understand the perspectives that are being shared, reflection of this understanding back to participants so that they feel heard, and synthesis and articulation of both legitimate differences and opportunities for common ground and consent (Schirch and Campt 2015).

CDSes like Polis and Remesh are often specifically designed to support certain aspects of this skill set, such as the identification of legitimate differences and points of common ground. Nevertheless, engagement outcomes are still highly dependent on the interactions between facilitators and participants, such as the prompts participants respond to, as well as the seed statements facilitators submit in order to help frame the dimensions of the discussion, and ensure that there are comments available to vote on for the very first participants who enter the conversation.

Some platforms, like Remesh, allow facilitators to pose a series of prompts for participants to respond to, enabling dynamic back and forth where the facilitator can adapt the conversation or probe a certain response to develop a more realistic, nuanced deliberation. Other platforms, like Polis, provide a single conversation prompt and rely on the real-time submission of seed statements as an alternative way of adapting the engagement to emergent topics within the context of one higher-level prompt[2].

We expect LLMs can play a significant role in reducing the effort and necessary expertise associated with such dialogue design and facilitation. An early example of this is Remesh's prompt suggestion feature which offers suggestions for rewording CDS prompts in order to mitigate framing bias and elicit quality responses. There remain, however, a wide range of opportunities to use LLMs to assist in dialogue design and moderation:
- **Translation:** LLMs enable text to be rapidly translated with high accuracy in near real-time into a large number of languages, expanding access to CDS for those who do not speak the dominant language. Multi-language deliberations and dialogues in the past were resource intensive, but LLMs enable CDS participation by a broader, multi-lingual pool of participants.
- **Intelligent dialogue generation tools (for facilitators)**: a prompt-tuned LLM chatbot can engage in a dialogue with a facilitator to co-design the CDS. The chatbot can pose a series of questions about the CDS a facilitator aims to run, then generate a dialogue design and seed prompts for the facilitator to use.
- **Intelligent prompting and scenario generation (for participants)**: Participants may not always provide rich statements in response to certain prompts because of a lack of understanding or preexisting opinions on the topic. This is especially true for complex policy topics or abstract questions about values. One mitigation here is to intelligently prompt participants to respond to LLM-generated scenarios. LLM-based tools can also improve the formulations of participants' statements by proposing reformulation or asking them for more context about their post. This can also operate at the level of the CDS, where an API can take the data collected by a CDS (statements + votes) and generate suggested follow up prompts or seed statements for a follow-up round of deliberation.  Here GATE and STaR-GATE (Andukuri et al. 2024), which are designed to ask clarifying questions of a single person, could be extended and generalized to engage with a wider participant body.
- **Bridging-based comment routing**: bridging systems can take an individual's response, as well as additional context like demographic data or voting history within a CDS, and produce a suggested follow up prompt or question specific to that individual, in order to gather additional detail on how they fit into the broader opinion landscape. These systems can also take the contents of an ongoing CDS and generate or elevate messages that garnered similar votes across opinion groups. This can enable participants to see and respond to shared common ground in real time.

---

[2] Occasionally, facilitators will create new conversations with more specific prompts to capture some of these dynamics. There are however drawbacks to this approach in that it becomes more challenging to see how responses compare across these separate conversations without custom analyses, as well as interface challenges for participants having to navigate between separate conversations.

- **Fact checking during deliberation**: many debates or deliberations today include professional fact-checking to provide near-real-time analysis of the veracity of claims made. AI could similarly help in providing accurate information in real time, ensuring participants operate from a shared set of facts (Landemore 2022a).

### Elicitation Inference

In nearly all uses of CDSes, there are far too many submitted position statements to expect each participant to vote on all of them. Consequently, most participants don't vote on most statements. In settings where there are no constraints on how many times a participant can vote on statements, large asymmetries in the number of votes cast by different sub-groups of participants can arise leading to biased vote tallies. One potential solution used by some platforms today is to predict how an individual would have voted based on the statements they did vote on. There are significant risks with this approach of simulating votes, though perhaps not as high as they are in vote simulations that assume a different level of knowledge than people actually have (Brennan and Landemore 2022). These need to be carefully designed for, including the perception of disenfranchisement or the manipulation of inputs through synthetic public input.

Such vote predictions are possible both because there are patterns in large-scale voting — the same type of patterns routinely exploited by recommender systems. There is also rich information in the semantic content of the comments themselves, which LLMs can unlock. Vote prediction in a CDS setting has been demonstrated using pure LLM prompting, and has been found to be well calibrated, but can be expensive to employ for larger data sets (Fish et al. 2023). Hybrid LLM-latent factor models have also been developed with significantly improved performance characteristics (Konya et al. 2022). There are a few key opportunities to build on this nascent work:

- Develop an open data set of CDS data from which elicitation inference performance benchmarks can be established and tracked for various methods.
- Create better approaches to elicitation inference that demonstrate higher accuracy than those available today, e.g. fine-tune an LLM to excel at this task.
- Solve the confidence problem, by developing an approach to elicitation inference which enables the margin of error to be estimated for results derived from aggregations which include inferred votes.
- Create an open toolbox for elicitation inference that inputs and outputs data in some universal format which enables interoperability between other tools, such as those for doing LLM based summarization and synthesis.

### Summarization and Visualization for Sensemaking

During and after a CDS, there are opportunities for producing easily digestible summaries of the voluminous data produced by these systems. There are, for example, Polis engagements which have involved tens of thousands of participants, submitting thousands of statements and millions of votes on those statements ("Aufstehen Case Study" 2018). Making sense of the output of these large-scale deliberations is a daunting task. Some CDS offer algorithms for mapping the opinion space, clustering participants into opinion groups, and ranking comments, all of which help with understanding what participants collectively said. Remesh provides a sophisticated interface for analyzing the voting patterns in terms of demographic data, and provides tools for summarizing and synthesizing results. Still, fully understanding the results and producing impactful outputs is currently an onerous, time-intensive undertaking, which often requires a meaningful amount of human labor. LLMs offer an opportunity to simplify these processes by augmenting human intelligence to more efficiently summarize and make sense of public opinion from CDS outputs.

Google DeepMind has shown that a fine-tuned set of language models can be combined with a preference aggregation framework to find common ground amongst individuals with diverse perspectives (Bakker et al. 2022). Building on this work, researchers found it possible to reach common ground on controversial topics in virtual deliberations through iterative participant feedback on LLM generated statements. Participants' initial statements can be synthesized with LLMs into "group statements" that aim to identify common ground. These group statements are then iteratively refined with participants' feedback and LLM synthesis to generate statements that maximize participant endorsement (Tessler et al. 2024).

Fish et al. proposed *generative social choice* as a framework for combining LLMs with social choice theory to generate a slate of statements that accurately represent a heterogeneous population (Fish et al. 2023). Meanwhile, the Computational Democracy Project, in collaboration with Anthropic, explored the application of LLMs to numerous challenges associated with CDS platforms, including the summarization of Polis conversation data (C. T. Small et al. 2023). Their work highlights the critical role of context-window length in producing nuanced and rich summaries of vote patterns across different opinion groups. Similarly, tools like Remesh and Talk to the City have integrated some degree of summarization and topic modeling to assist with the analysis process (Gallagher 2024). These developments collectively point towards a future where AI can significantly augment facilitators' ability to understand and synthesize complex, large-scale conversations, affording the opportunity to significantly enhance the impact of CDSes with results that are more easily consumed.

This work has obvious advantages for policy makers. Martin Hilbert described a "Deep Democratic Neural Network", a system whereby an AI would summarize citizens' political preferences, input via written text statements, into a single policy platform to be acted upon by politicians, now equipped to better represent their constituents on more rich information than a single vote cast every few years (Hilbert 2009). 15 years later, the technology now exists to enable the collection, synthesis, and visualization of citizens' nuanced political perspectives for decision makers of all varieties.

Early progress to this end is underway with Policy Synth, a modular system designed to synthesize the findings from engagements together with research findings to create evidence-based policies. For example, New Jersey's AI Task Force is using Policy Synth with the platform All Our Ideas to gather input from workers on what they see as the greatest opportunities and challenges regarding the impact of generative AI on the state's workforce. The output from the engagement will be a rank-ordered list of public concerns for policy makers (Gambrell 2024).

## Platform for Integrating Deliberative Systems

Presently, stitching together meaningfully deliberative processes out of these tools is not straightforward. There is an ecosystem of technology that can facilitate public engagement, but very little attention has been given to making these pieces composable. Progress towards this goal requires coordination amongst the various platform creators, maintainers and practitioners to hone in on appropriate usage patterns, data formats and protocols.

Many teams using CDSes consistently produce similar reporting across multiple projects and have common reporting outputs, such as percentage agree or disagree with a statement, or number of new statements added. There is a clear opportunity to apply the principles of reproducible analytical pipelines ("Reproducible Analytical Pipelines (RAP)," n.d.) and the associated software engineering practices to make such reporting more reproducible and less error-prone.

Ideally, a meta-platform for CDS would:
- Support composition and interoperability with modular design, empowering process designers to adapt their methods to different circumstances;
- Offer core infrastructure that these platforms can build on, accelerating innovation of new technologies in this space, and improving interoperability of existing technologies;
- Provide a sandbox for testing tools, including with simulation, so that practitioners can experiment and gain intuition as they design improved tools;
- Come with built-in evaluation and benchmarking infrastructure for tools and processes;
- Support end-to-end CDS workflows from recruiting to report generation.

### 1.3 Risks and Mitigations for AI-enhanced Collective Dialogue Systems

While CDS and deliberative technology stands to benefit greatly from application of LLMs and related technical innovations, this adoption does not come without risks:

- **Aggregation of opinions is one part of deliberation:** It is important to distinguish the use of LLMs in CDS to solicit high-definition snapshots of public opinion from mass deliberation, the latter of which involves an iterative exchange of reasons, arguments, and justifications (Landemore 2022b). LLMs can help scale such deliberations, but for the context of mass deliberations, LLMs should be understood as augmenting the process rather than substituting wholesale.
- **Misrepresentation owing to use of LLMs**: LLMs can occasionally hallucinate information and struggle to represent the opinions of minority groups (Agnew et al. 2024). Additionally, if LLM-based simulations of deliberations are used as part of a system, there is a further challenge of assessing the veridicality of the outputs. For example, designers will need to work to ensure the LLM equally or proportionally represents different viewpoints in its syntheses, rather than over-emphasizing certain viewpoints.
- **Appeals & Ethical Review Mechanisms**: It is possible that LLM summaries or participant inputs may include offensive or disputed content that are not moderated by CDS facilitators. For such dialogues to be inclusive and responsive, it is important to include appeals processes for participants to report issues and trigger a procedural investigation into breaches of a pre-existing policy on norms of the dialogue. Such an appeals body could proactively review the LLM-generated content before publication, in addition to participant appeals. This body would need to be independent from CDS facilitators (Landemore 2022b). This risk is further mitigated by publishing moderation guidelines and norms publicly ahead of any CDS and including these in participant education prior to the CDS. Transparent access by the public to all CDS participant statements, including those moderated out, can also help mitigate ethical concerns.
- **Synthetic participation is promising yet democratically problematic and potentially risky:** LLMs can already participate in CDS and other forms of deliberation in place of, or in addition to, human participants. This approach offers the potential of more representative results (by projecting votes beyond what human limits of time and attention are capable of giving voice to). But it runs the risk of degrading the CDS' legitimacy, diluting the public's agency, and misrepresenting reality in convincing ways. The transparency of such synthetic vote projections is paramount to avoid misperceptions of AI replacing humans or manipulating outcomes. Can the advantages afforded by synthetic participation be realized while preserving democratic legitimacy and safely managing the risks?

## 1.4 Future Research on AI-enhanced Collective Dialogue Systems

- **Design for "deliberation in the wild" on existing platforms:** The existing digital public squares can be designed to promote healthier discourse and mutual understanding. A core area of future research could focus on prosocial designs on these more engagement-based platforms to incentivize collective dialogue.
- **Assess trade-offs imposed by current CDS design constraints:** CDS systems allow participants to submit comments and vote on others, but not engage in threaded discussions. Many virtual CDS enable anonymity, use text only inputs as opposed to richer formats (e.g. video, audio, images), and lack synchronicity. Further research is needed to understand the impacts of these design constraints, and whether CDS outcomes can be improved with different normative modalities.
- **Understand the cost of inaction:** It is difficult to quantify the benefits of embracing deliberative and participatory technologies. Further research is needed into measurable outcomes, both qualitative and quantitative, between participatory processes and the counterfactual, non-participatory case.
- **Motivate healthy participation:** A successful CDS requires sufficient and representative participation. However, the process of effectively recruiting and motivating participants is unclear in most contexts. Paying participants motivates them to participate, but may crowd out other incentives (sense of civic duty, intrinsic reward of participating) and be unsustainable at scale. Future research should not only aim to understand appropriate motivation strategies, but also take seriously the lack of interest in participating expressed from many, and aim to address this participant disinterest or apathy.
- **Build for accountability and transparency to ensure trust in AI and the CDS:** Public excitement about AI is currently mixed together with skepticism and uncertainty. The successful deployment of AI tools to benefit deliberation and collective dialogue will likely hinge on our ability to create credible accountability mechanisms, ensure process transparency, and understand participants' shifting mental models of AI and disclosure and transparency needs as well as the drivers of and barriers to trust in AI in specific circumstances. Successful CDSes require both process legitimacy and output legitimacy, and trust in the underlying technology could significantly affect this perceived legitimacy.

## 2. Bridging Systems

Most online dialogue about substantive issues takes place on consumer platforms that were designed more for entertainment and attention farming than for the discussion of contentious social issues. How might these more widely used systems benefit from some of the same mechanisms that power collective dialogue systems?

One idea is that our everyday systems should be designed to encourage trust, understanding, and building of common grounds between people with different views, values, and identities. This concept is commonly called "bridging" (Ovadya and Thorburn 2023). Building trust in divided societies is part science and part art, with specialists focused primarily on complex, time-intensive efforts to enable awareness of shared values and interests. Dialogue, negotiation and mediation processes have long used the concept of bridging to advocate for "win-win" solutions that address the underlying interests and needs of key stakeholders (Schirch 2024). Advances in AI language capabilities are expanding possibilities to scale bridging beyond time intensive in-person interactions.

Existing content selection algorithms sometimes create more division. Most of the algorithms that personalize content for us – often called recommender systems – try to select those items that we will click on, share, or otherwise engage with (Thorburn 2022). Engagement benefits businesses as the more a user engages with content, the more ads can be shown and the more content is generated to attract other users.  However, there are times when the most engaging content is also the most polarizing (Rathje, Van Bavel, and van der Linden 2021), especially since engagement is often dominated by small groups of more abusive users (Hindman, Lubin, and Davis 2022).  This can have a variety of negative consequences for how people relate to each other, including but not limited to trapping people in echo chambers, incentivizing publishers to share more divisive content, and contributing to individual radicalization towards violence or societal escalation towards conflict (Stray, Iyer, and Larrauri 2023).

## 2.1 Current Applications and Challenges of Bridging Systems

Aggregating signals from diverse users has been a part of tech products for decades. Internet sites like Metacritic and Ranker have built popular consumer platforms by aggregating diverse opinions into crowdsourced information. Google's pagerank algorithm uses links from diverse webpages as a signal of trust and Meta has used a similar signal within its systems to fight misinformation (Rodriguez 2019).  All of these systems require some understanding of the diversity of the signal, in order to reduce the correlated error that comes from spam, manipulation, or a particularly active group of users.

In the last few years there has been a more specific focus on bridging signals, defined in this context as where there is agreement or overlap across diverse users. Meta incorporated "diverse engagement" into comment ranking in 2021 after internal experiments found that prioritizing comments that receive positive engagement from diverse other users led to quality improvements such as "less polarizing content," decreased views of "bullying comments," and fewer violations of community standards. Comments on civic or political topics with "diverse positive engagement" were perceived by users as "healthier" and "good for the community."  There were also gains from upranking content from users with a history of writing posts that garnered diverse approval. These changes also led to a 0.69% increase in Facebook comment views, a significant number in the context of platform optimization ("Fbarchive.org," n.d.).

In Meta's work, diversity was calculated using the World2Vec vectors of individual users. World2Vec is a trained model in wide use at Facebook that produces an "embedding" for each user, an opaque sequence of numbers that summarizes the character of their social network and interaction history. These vectors can be used to calculate a similarity score between any two people, even though they do not directly encode sensitive information like demographics or political views (Lerer et al. 2019).

X's Community Notes allows users to propose labels or "notes" to posts that might be misleading. Only notes that are endorsed by diverse users are displayed (Wojcik et al. 2022). Inspired by the algorithms used by Collective Dialogue Systems like Polis to identify consensus, Community Notes works by separating a user's rating of a note into two contributing factors: how much that user agrees with the politics of the note, and how broadly helpful that note is independent of politics. Those notes which score highly along the helpfulness dimension are selected for display on the corresponding posts.

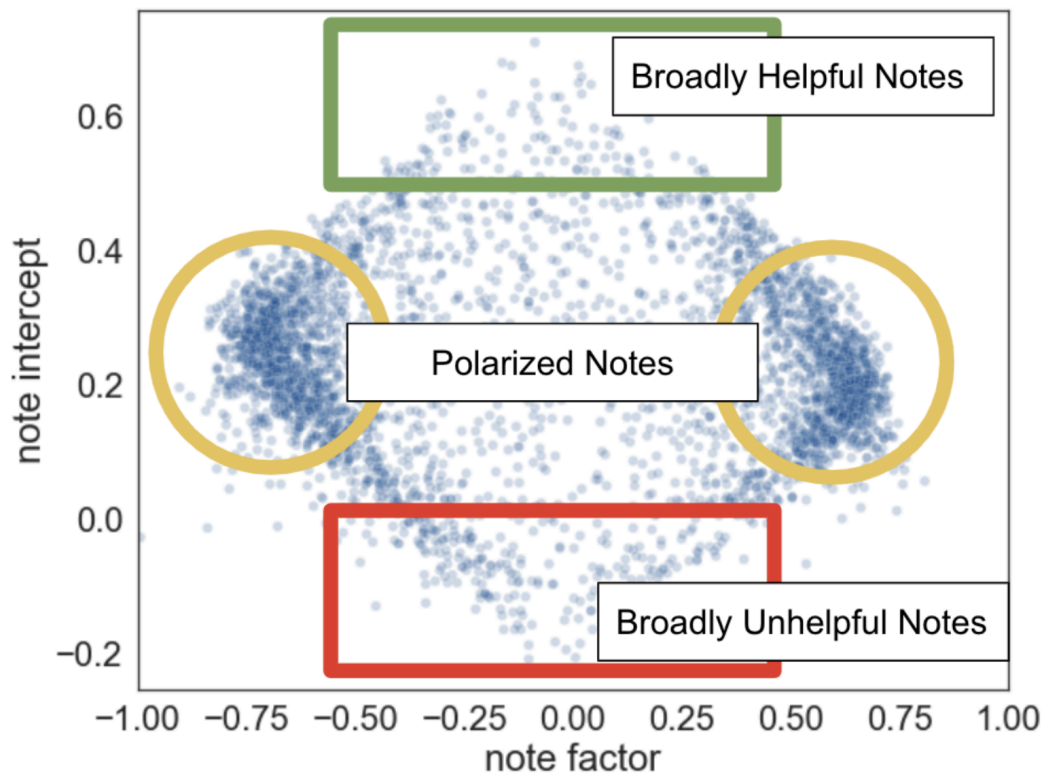

Figure 3: This diagram illustrates how the X Community Notes algorithm works. Note ratings are decomposed into two factors: political orientation (here, the horizontal axis) and helpfulness independent of politics (vertical axis). Notes that score low in partisanship and high in helpfulness are chosen for display (Wojcik et al. 2022).

This is a matrix factorization approach, of the kind commonly used in recommender systems to predict which items a user will like (Koren, Bell, and Volinsky 2009). In this case, exactly one factor is calculated for each note and each user, representing ideology or perspective along one axis. This axis isn't necessarily left-right, nor is the algorithm trained on any sort of political data set. Rather, this method finds the primary axis of division between users, whatever that axis may be in the context of those specific users and posts.

In practice this approach selects misinformation labels that are both more persuasive and more widely perceived as helpful. It beats other algorithms such as supermajority voting or simple average rating (Wojcik et al. 2022).

YouTube also released a feature allowing the crowd to add notes to videos leveraging bridging. Although they have not published the rating algorithm, it is "a bridging-based algorithm" that "helps identify notes that are helpful to a broad audience across perspectives" ("Testing New Ways to Offer Viewers More Context and Information on Videos" 2024).

Bridging can also be based on content analysis, which requires no information about the user and requires fewer inputs, meaning it may simplify explainability and implementation. Jigsaw's Perspective API has leaned into this content-based approach to bridging by adding signals for attributes users would likely value in conversation such as curiosity, nuance, personal stories, etc. using a suite of trained models ("Perspective API Bridging Attributes," n.d.). This would enable comment hosting platforms to be consistent and explicit about their guidelines for participant interaction as well as how they score and rank posts. Another benefit here is that platforms and moderators wouldn't need to collect additional user information or have prior comment platform activity history; a win for privacy and moderators of small communities who might not have platform activity access. Additionally, identifying explicit qualities of bridging as opposed to leaning on a black box of agreement across parties could help with fairness and algorithmic explainability. It is likely that some combination of both methods (user diversity and content quality) will be necessary for optimal bridging outcomes.

The above examples are just a sample of what could be possible by adding bridging signals to systems. While current bridging implementations have focused on finding consensus content, addressing information quality, or reducing bad experiences, there may be demand for systems that connect diverse people or increase empathy.

## 2.2 Bridging Signals

As the examples above show, there are different ways to think about what "bridging" means, and correspondingly many metrics or scores or indicators that might indicate which content has high "bridgingness."

Many bridging systems are based on some numeric measure of similarity or difference between users. Once there is a way to know that two users differ along some axis, many kinds of bridging signals are possible. For example, the idea of "diverse engagement" is to select content that diverse users would "like" or engage with in some positive way.

It is possible to identify users as similar or different based on the content they post or consume. There are a variety of techniques for turning collections of content into abstract feature vectors or embeddings that can be compared to measure the degree of difference. One example is topic modeling, which can be used to determine if two texts are on similar subjects. However such general content analysis methods might differentiate users because they post on different topics (e.g. beekeeping vs. Polish cinema) rather than having opposing positions on the same topic.

Alternatively, there is increasing use of models trained to capture the aspects of individual posts that are relevant to bridging. This line of work began with evaluating sentiment, emotion or incivility, but has progressed to more complex constructs such as "conversational receptiveness," a score that predicts positive discussions (Yeomans et al. 2020). The experimental "bridging" scores recently added to Jigsaw's Perspective API evaluate comments for affinity, compassion, curiosity, personal story, or reasoned argument ("Perspective API Bridging Attributes," n.d.).

Behavior-based signals use clusters of behavior to define distance. Users who follow different influencers, buy different things, or use different site functionality will be scored as more distant than users who have more in common behaviorally. Behavioral methods can be truly content agnostic, in which case they are likely to pick up on geographic and language differences. Or they can be narrowed to more socially relevant domains, such as whether users follow differing political news sources (Cunningham et al. 2024).

Another approach is social network analysis, where patterns of friending, following, or other interaction reveal clusters of like minded people (Conover et al. 2011). Generally, people associate more with like minded individuals, and in highly divisive or polarized settings they tend to avoid or even cut off contact with people from "the other side." This means that clusters in the social network often correspond to particular sets of values and attitudes. Alternatively, users might simply be asked to self-identify along relevant axes of difference. This is likely to produce accurate data, but comes with many concerns about appropriateness, privacy, transparency, and misuse (Andrus et al. 2021).

Each of these methods for identifying dissimilar users captures a different type of diversity: what they say, who they interact with, how they identify. Behavioral and network approaches are likely to be more palatable for companies that do not want to collect or infer data with political, demographic, or religious ramifications, especially given legislation that requires the disclosure of such inferences when asked by the user (such as the GDPR).

For many methods there is likewise no need to try to describe racial, political, or ideological divisions, such as providing training data describing what "left" and "right" mean. Instead these methods try to find the primary axes of division, the main ways that users differ and disagree. However, this broader approach may mean that bridging systems fail to specifically address the most important divisive issues of our time. For such cases, it may be more beneficial for platforms to create dedicated spaces where self-identified groups can engage in meaningful dialogue across differences, or in collective dialogues.

## 2.3 Opportunities with Bridging Systems

These metrics used to select content are conceptually different from metrics used to evaluate the success and future opportunities with bridging. There are different reasons for using bridging systems, leading to different ways of measuring their success.

In the case of Facebook's "diverse engagement," success was measured in terms of existing quality and integrity metrics. Including this signal led to fewer views of content rated low quality by existing integrity metrics, and fewer user reports of content violating platform rules.

Bridging might also be done with the goal of improving relations between groups, such as increasing understanding or reducing polarization. For these broader outcomes, social scientists and conflict professionals typically use surveys asking users about how they perceive the outgroup, that is, people on the "other" side of some division. Examples include affective polarization, meaning how warm or cold someone feels towards the outgroup, and various scales of distrust and dehumanization (Finkel et al. 2020). Surveys provide a type of ground truth that may not exist in platform behavioral data.

It is relatively difficult to algorithmically influence broad user attitudes (Cunningham et al. 2024) – for one thing, much else is happening in someone's life – so the challenge is to find a survey question that is narrow enough to move yet broad enough to be relevant. For example, general well-being measures seldom change during platform experiments, but questions like "have you had a negative experience on this platform in the last week?" can be much more informative.

Sometimes it is possible to train a model to predict the survey outcomes of displaying certain content. When this becomes reasonably accurate such a predictive model can be used to help identify appropriately bridging content, and many platforms optimize for survey outcomes in this way (Cunningham et al. 2024). Survey data is still quite scarce and expensive. Ideally, the outcome of bridging could be assessed from the purely behavioral user data that comes from regular platform use. There are several types of plausible behavioral signals, both positive and negative:

- How often people engage positively and negatively with outgroup content
- How often people seek negative outgroup content, such as through search or subscriptions
- How often people post content that scores high on bridging signals
- How connected or divided the social network of two opposing groups is
- Fraction of content that is seen and positively interacted with by both groups, akin to the "average content bridgingness"

Any of these could be used for evaluating the effectiveness of a bridging system, especially if their relationship to traditional survey measures can be better understood.

Online public squares could instead be designed to bridge political divisions. One problem with homogenous online discourse is that it allows for more division and misinformation, as there is less room for diverse audiences to correct correlated error (Shi et al. 2019). However, simply increasing the heterogeneity of online content is not always received well; exposure to uncivil opposing content, compared with exposure to civil disagreeing content, can lead to greater levels of attitude polarization (Stray 2022). The design of digital public squares can amplify or diminish political divisions.

To help facilitate bridging across these divisions, we can design spaces civility in public discourse, and leverage LLMs to help identify and amplify the rhetorical strategies that promote trust and mutual understanding. Examples of these include constructiveness, sharing personal stories, and proposing solutions (Stray 2022; Kolhatkar et al. 2020). Research has also shown that upranking specific qualities of conversation such as constructiveness and curiosity can increase understanding and reduce partisan animosity (Saltz, Jalan, and Acosta 2024). There is a reasonable argument that such changes to ranking systems could be beneficial for business goals given that consumers say that toxic political discourse leads them to use products less and some platforms have found that improving the quality of content can lead to long term usage gains (Cunningham et al. 2024).

## 2.4 Risks and Mitigations of AI-Enabled Bridging Systems

Rewarding bridging content may sometimes be at odds with constructive conflict and disagreement that are essential to democracy. Increasing mutual understanding might not always be the main objective of a community, especially where it is important to express dissent or protest. Bridging could lead to a decrease in trust in the platform or community, if these systems are perceived to be illegitimate or opaque. Given these risks, it is important to architect and implement bridging systems with transparency and a diversity of signals for ranking (Ovadya and Thorburn 2023).

One risk of bridging systems that leverage LLMs is the elevation of generic, broadly appealing but insubstantial content at the expense of meaningful discussion. Bridging changes ranking incentives and can thus introduce the risk of users gaming ranking algorithms, posting content explicitly for the purpose of being upranked. For example, if a platform explicitly upranks curiosity and compassion, users could game the system by posting sarcastic comments with ironic curiosity or pairs of empathetic responses to cruel remarks. Beyond ranking abuses, there are important privacy considerations related to classifying a users' behavior or content as more or less bridging. This is akin to the existing privacy issues that arise with predicting which content is engaging.

## 2.5 Future Research on Bridging Systems

Bridging systems have great potential to improve mass market discussions of important topics. We have powerful examples that illustrate the benefits as well as the way we would implement bridging and measure those benefits. However, several open questions remain:

- *Can bridging be gamified?* Companies have gone to great lengths to gamify engagement and could find ways to reward users for writing bridging content. Some example implementations could be amplifying their content, giving them badges or points (e.g. Reddit karma), and encouraging comedy. Could we trigger interventions by people who opt-in to be bridged, via survey questions?
- *Could we incentivize bridging amongst producers of online content?* Experiments could also be done on producer incentives. Bridging algorithms should not just change the order of the content that already exists, but also also change the incentives for what kind of content gets produced.
- *Could UI elements encourage bridging?* Platforms could also experiment with helping users to connect on dimensions of similarity. For example, when users share information, platforms can highlight common identities or interests such as, "you are both parents" or "you both like classic cars". Platforms can also highlight when people take shared actions to reinforce common ground and empathy, such as clicking a "respect" or "like" button, or supporting the same social movement (Stroud, Muddiman, and Scacco 2016).
- *How can bridging experiments be done more outside of large companies?* Ideally we could test ideas that do not require permission from social network platforms. One common approach is using a Chrome browser extension to modify a platform's UX. Alternatively, researchers can test bridging in synthetic social media experimental sandboxes, though these have challenges with ecological validity.
- *How does bridging affect long and short-term engagement?* Given that some quality interventions have improved user retention metrics (Cunningham et al. 2024), under what conditions and time horizons can we conclusively say that bridging can improve engagement?
- *How can we build bridging systems that are trusted and seen as legitimate by people?* Explainability, transparency, and control, including responsiveness to user input, will be key levers for ensuring trust and legitimacy of these tools. Given existing popular perceptions about platform neutrality, as well as public skepticism about whose interests AI innovation is primarily serving, how can we ensure that bridging technology is seen by people to serve their interests?

## 3. Community-driven Moderation

Even if online systems are designed with bridging in mind, some amount of moderation will always be necessary. Yet AI offers potential to ease the challenges of moderation where human moderators are overburdened, while giving more control and context to the communities themselves when it matters to them most. This section will discuss how putting moderation in the hands of communities, rather than platforms, can enable this process to be more robust and legitimate.

Many user-generated content (UGC) platforms like Reddit, Discord, and Facebook Groups have two tiers of moderation: "commercial moderation" for platform-level or application-wide policies, typically led by paid reviewers using dedicated moderation tools, and "community moderation," led by volunteer community members engaged in a specific topic or groups on a platform. The practice of community-centric moderation can be further distinguished from crowdsourced moderation on platforms like Wikipedia, Google Maps, and X's Community Notes, where all end-users can take specific moderation actions, such as editing or correcting an entry according to platform-wide policies, such as factual accuracy. While there are many models of governance for the "middle level" between platform representatives and end-users, as described by Jhaver and co-authors, we focus here on community moderation within groups (Jhaver, Frey, and Zhang 2023).

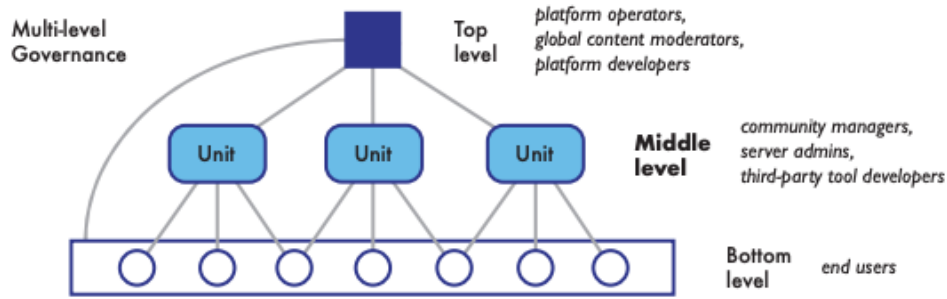

Figure 4: A simplified illustration from Jhaver, Frey, and Zhang of multi-level governance of an online platform. This paper focuses on middle-level units, which comprise community managers and moderators, server administrators, third-party moderation tool developers (Jhaver, Frey, Zhang 2023).

## 3.1 Current Applications and Challenges with Community-driven Moderation

Community moderators develop, refine and uphold criteria for acceptable engagement in community spaces by determining what are relevant practices. For example, a group for sharing memes and humor may choose to disallow posting knock knock jokes, or a disability group may disallow unsolicited advice. Members of these groups do not necessarily expect these rules to be upheld across a platform, but find value in upholding them within their group to maintain trust and engagement. This is because norms and values are contextual. A work meeting and a bachelor party may take place in the same city, with the same laws, but each has different standards for acceptable social behavior. In some cases a community's rules may even seem to be in tension with the platform's broader policies; for example, the Reddit community, r/BlackPeopleTwitter, includes racial verification as a moderation practice — a policy that could be considered exclusionary at a platform level (Smith et al. 2024).

### Lack of Platform Support

Training programs and explicit support from platforms like FB Groups, Reddit, and Discord have declined over the years, and in some cases have devolved into outright hostility between platforms and moderators. Discord initially invested heavily in moderator upskilling, including training, exams, tooling, and even social norms incentivizing the programs like badges. However, Discord discontinued its Moderator Academy program in 2022 (*A Discord Moderator's Worst Nightmare* 2022). Reddit shut off free access to its API in 2023, effectively shutting down most third-party Reddit clients, many of which were used by moderators as they included tools not available in the official Reddit client. Reddit users protested site-wide, and many subreddits went private. Moderators of large subreddits who refused to return their communities back to public were replaced with moderators chosen by Reddit (Peters 2023). Stack Overflow moderators went on strike in August 2023 over the site's decision to prohibit the removal of AI-generated content. After negotiations between the company and representatives of the striking moderators, Stack Overflow agreed to allow removal of AI-generated posts ("Moderation Strike: Results of Negotiations" 2023).

In other cases, companies may be interested in using user moderation data to train LLMs to replace commercial human moderators or volunteer community moderators. While this may work in some cases, in others it is wildly inaccurate – one study found that ChatGPT 3.5 performed worse than a coin toss when moderating subreddits that required significant expertise and context to evaluate content (Kumar, AbuHashem, and Durumeric 2024).

The tools provided by platforms are often inadequate to the complex needs of moderators. For example, many community moderators are faced with large quantities of AI-generated "slop," which is swamping moderation queues and decreasing the utility of their community. Facebook Group moderators are forced to manually evaluate "join requests" of potential new members, which often requires complicated judgment calls to evaluate the authenticity of the account (Kuo, Hernani, and Grossklags 2023). Moderators need tools to handle raids, generative AI, spam, and karma farming, among other challenges.

### Volunteer Burnout

Moderators face burnout due to the volume of content, being targeted by users for their decisions, the emotional labor required for moderation, the lack of specialized tools and training provided, and exposure to offensive content (Schöpke-Gonzalez et al. 2022).

First, the workload of moderation can be considerable and often involves both visible and invisible labor (N. J. Matias 2019)(Li, Hecht, and Chancellor 2022). As a result, many moderators prefer to work in teams to split the workload with others (Gibson 2023). However, in practice, some moderators end up doing more of the work than others; one study found that "too little available time" and "struggles with other moderators in the group" were the two main reasons why volunteer content moderators quit (Schöpke-Gonzalez et al. 2022). If nobody else steps up to moderate, the quality of the community may go down; in some cases, moderators choose to close the community rather than enabling someone else to moderate it.

Second, many users fail to appreciate the benefits of moderation, instead seeing moderators as fickle, sometimes even oppressive or antagonistic, overseers. One study found that Reddit moderators often received complaints about their moderation decisions, which frequently included "profanity, racist slurs, and threats of violence" (N. J. Matias 2019). Third, to our knowledge, platforms rarely provide tools or formal procedures for mediation, which can be an important way to de-escalate and resolve conflict while maintaining community norms and ties.

### Quality control and problematic moderators and processes

The selection of moderators and the implementation of human rights safeguards in their decision-making processes are crucial issues in online communities. For instance, in scenarios where racist groups expel members for counterspeech, the need for fair and equitable moderation is evident. The quality of moderation can vary greatly, with some moderators potentially veering off course, which raises concerns about consistency and accountability in their roles.

Additionally, there are communities with norms that might be considered anti-democratic or harmful. Moderators may act according to community standards, but these standards may conflict with platform or social norms, as in the case of QAnon or anti-vaccination communities. Cultural differences can lead to behaviors that are normative in one context being problematic in others, such as insider language, comfort with nudity, or religious language (Kaye 2019).

In other cases, communities may have conflicting internal norms. In 2020, when concerns around racial justice in the United States were high, several prominent Facebook "mom groups" splintered between women who did and did not want to talk about political issues (Lorenz 2020). In such cases it is difficult to determine who is "right" and "wrong" without using normative judgment. Moderators may not always know what the "right thing to do" is, or might not know what their communities actually want.

There is a need for transparent mechanisms for making moderation decisions. Internally, platforms often obscure their content moderation guidelines in order to avoid users "gaming" them, but this is especially problematic in self-governed communities where moderation runs on trust and community buy-in. How do moderators use platform policies and legal frameworks to guide their decisions and ensure they are just and transparent? This is most apparent in the process for appealing moderation decisions: users whose content is removed or accounts banned are often angry or frustrated, as discussed above. Users may be dissatisfied with a moderator's stated reasons for removing their content, or they may not get feedback at all. There is rarely explicit support for sanctioned users regaining good standing within the community. Across a variety of platforms and communities studied, moderators and community members need better communication and support systems to understand and navigate moderation processes. What role do community members have in moderation, and how can — and can't — AI support them?

## 3.2 Opportunities for AI-enhanced Community Moderation

Some of the challenges that community moderators and members face could potentially be addressed or alleviated with AI-powered tools, including both tools that moderators can use to proactively maintain a healthy and prosocial environment, and tools that can support moderators to reactively deal with issues after they have happened.

### Proactive / Preventative Tools

While community moderation is often thought of as solely taking down content or removing accounts after a violation, moderators can also proactively set the norms of their community to encourage voluntary compliance and broadly shape the discourse and behavior in their community.

- *Moderator Listening and Understanding:* AI tools are adept at analyzing and summarizing large quantities of community interactions, which could be used to support moderators to quickly "take the temperature" of what has happened in their community. For instance, this could work similarly to AI generated summaries of reviews on Amazon. With these tools, moderators could better understand how their moderation work is perceived, enabling a positive feedback loop (Weld et al. 2024). Predictive models could also help identify potential viral events or spikes in moderation needs, allowing for better resource allocation and preparation.
- *LLM Simulations***:** Particularly for newer communities who do not have many members or content yet or newer moderators, moderators may not have a good sense of what policies to set for the community or what might result from taking a particular action. In this case, LLMs can simulate comment threads to prepare moderators for potential activity that may occur and allow them to explore different scenarios where they intervene (Park et al. 2022).
- *Community Guideline Creation:* LLMs could not only help moderators define policies and guidelines for their community but also involve community members or teams of moderators in co-defining shared values and norms (C. T. Small et al. 2023). This could be done through LLM analysis of community-generated content as a starting point, or as support during group brainstorming and visualization exercises. Indeed, involving communities in the development of moderation norms and practices could lead to more effective and accepted rules.
- *Author Feedback and Community Nudges:* AI could be leveraged to help would-be posters, either mods or community members, (Wright et al. 2021) before posting . Such feedback could teach people how to rephrase their writing to convey a different tone (e.g., be more polite, empathetic), adhere to certain standards of writing (e.g., adding references), or otherwise stay on topic and adhere to community rules. Prior work on nudging users by reminding them of community rules has shown it can be effective in promoting healthier online interactions (J. Nathan Matias 2019) and reducing toxicity in comment spaces (Jigsaw 2020).
- *Conversational Bots Engaging in Community Conversation*: Not only could AI tools help people before they make a post, AI conversational bots that play a social role in community conversations could be used to help with mod tasks that lead to burnout and have difficulty scaling, e.g., rule reminders, performing de-escalation, other forms of discussion facilitation (Seering et al. 2018). Bots have been instrumental on platforms like Discord and Reddit and communities like Wikipedia (Zheng et al. 2019). For instance, automated explanations of post removals can improve subsequent user compliance (Jhaver, Bruckman, and Gilbert 2019). However, such bots should be labeled to make clear to community members they are interacting with a bot.

### Reactive Tools

There is great opportunity for AI to help alleviate the burdens of reactive moderation tasks that mods must perform.
- *Steerable / Customizable Moderation:* Mods often use rule-based tools to automatically flag, categorize, and prioritize comments. These include simple word-based filters (Jhaver et al. 2022) to more complex regular expressions such as Reddit's Automoderator "AutoMod" (Jhaver et al. 2019). In some cases, platforms like Twitch provide AI models that mods can individually tune. New AI techniques could help mods with authoring more sophisticated automated filters customized to their needs that work better than current approaches without requiring technical expertise (Wang et al. 2024).
- *Triaging.* Moderation work involves interfacing with community members who are submitting reports and appeals, sometimes in collaboration with other moderators, to make decisions on difficult cases. AI tools could potentially help with triaging these tasks, including helping to compile evidence for reports or other corroboration, prioritizing high-importance reports, delegating work between moderators, or flagging problematic misuse of reporting systems (Crawford and Gillespie 2014). For example, AppealMod provides automated support for users appealing violations and was shown to help with mod burnout (Crawford and Gillespie 2014; Atreja et al. 2024).
- *Post-Moderation Governance and Assessment.* Some communities have experimented with more complex governance procedures for giving out sanctions and restoring users after a violation. For instance, researchers have proposed and some communities have tried graduated responses such as "three-strike" systems, graduated sanctions such as account suspensions, jury trials (Kou et al. 2017), or restorative justice processes. Lack of wider adoption of more complex governance procedures may partly be due to the challenge of designing such software infrastructure or the ongoing monitoring and assessment required, both of which could potentially be alleviated using novel AI tools. The PolicyKit for Building Governance in Online Communities published describes a range of governance models such as a random jury deliberation, reputation systems, and a promotion procedure inspired by Wikipedia's Request for Adminship process that can maintain community harmony through complex moderation decisions (Zhang, Hugh, and Bernstein 2020)

## 3.3 Risks and Mitigations for AI-enhanced Community Moderation

While AI has significant potential to help with community-driven moderation, it also has risks and limitations that may hinder the legitimacy and effectiveness for the communities it is meant to serve. For example, opportunities to augment moderation with AI through user-facing *Authorship Feedback and Community Nudges* may work well in certain contexts, but can also be biased towards certain styles of language that are better represented in LLMs, subsequently disadvantaging underrepresented groups. In addition, as described above under *Quality control and problematic moderators and processes,* even aligning on community norms in the first place may be a fraught process that can lead to disagreements and conflicts, especially in communities with diverse values. It is crucial to ensure fairness and inclusivity in the norm-creation process, so that these diverse voices are heard without violating the rights of others. More generally, AI-supported community moderation technologies should be designed with sociotechnical drivers of user trust and uptake in mind, if they are to be accepted and taken up by communities that might benefit.

There are also things that AI can not, or should not, be expected to address without human input, such as relational conflicts, crisis management, and economic incentives for moderation. To mitigate some of these challenges, we recommend the following:

- *Pay the Moderators or Invest in Community Leaders:* Provide financial or other incentives for engaged moderators to fairly and equitably compensate them for their labor, and ensure systems are addressing real problems faced by communities. Even as tooling improves, the work of relationship building and dialogue is at its core a human enterprise. Even if rewards aren't financial, more support by platforms can help reduce burnout and empower them to do the community work that matters to them. Fundamentally, this requires platforms to invest in co-creation processes and support community-driven moderation efforts.
- *Support Moderators in Crisis Management:* No matter how far tooling comes in reducing manual labor of moderation, community moderators will inevitably face crises within their spaces, such as threats of self-harm and suicide, that are most humanely and ethically addressed through human connection and professional crisis management. Platforms can provide training on crisis management, or even offer specialized tools that can help them better cope with their challenging roles. AI may assist, but ultimately should never replace, such support.
- *Keep Human Moderators and Communities in the Loop:* As advanced as automated flagging systems may get, language is dynamic and contextual and therefore it will always be valuable to have input from the people embedded within the community. Even as AI reduces the moderation workload, shifting human moderators' role toward quality control, AI developments in moderation should not be thought of as wholesale replacement for nuanced, complex community moderation efforts.

### 3.4 Future Research on Community-Driven Moderation

Future research depends on the development of better metrics for evaluating the success of community-driven moderation. Given that standard benchmark datasets for AI are not representative of the language of specific spaces, typical evaluation metrics like AUC - ROC say little about the actual quality of LLMs as applied for a given community (Narkhede 2018). Further, those users and moderators judge effectiveness in ways that may be less readily quantifiable, such as the mental toll of moderation or social cohesion within a group. More foundational research is needed to define which outcomes and metrics can be used to understand how different moderation solutions may be helping or hurting communities. Researchers should also investigate how these outcomes and metrics may vary depending on the type of platform and community.

With these outcomes and metrics in mind, both platform and independent researchers can design and study effects across the range of proactive and reactive opportunities described above, and more. For example, in proactive opportunities, researchers could study the efficacy of nudges that reinforce or build on existing community norms. Research could also investigate the potential role of the source of the nudge (for example, from a creator or bot) and contexts (1:1 or in a group space) for different niche communities. For reactive moderation, there is abundant research to be done on the efficacy of moderation based on content identified by steerable LLMs as compared to existing systems, as well as on the efficacy of AI-powered systems for designing and enforcing alternative governance structures, like juries, and in streamlining appeals.

This work could also be advanced by further research into present and historical patterns around trust and legitimacy (how it is made and unmade) within online community spaces, especially as AutoMod tools have proliferated and automatic or algorithmic judgments have been playing a greater role in moderation. Understanding these dynamics will be crucial for designing more advanced AI interventions that can run with the grain of user trust criteria in these communities, not against it.

Finally, an important area of study is to understand the potential of layering techniques. For example, community-driven moderation may also be enriched by approaches described in previous sections on bridging systems and deliberation and collective dialogue processes. More research is needed to understand the uses of deliberative conversations by and for online communities, specifically, and how these may aid in the development of norms, rules, and other practices typically associated with community moderation. While community-driven moderation may increase perceived legitimacy of moderation within communities, there remain distinct risks associated with both identifying and anonymizing community members involved in those processes.

## 4. Proof of Humanity Systems

One issue with digital spaces is that they can become dominated by small groups of highly motivated partisans, who are willing to mislead and attack others in the service of their aims (Hindman, Lubin, and Davis 2022). As AI technologies evolve, it will be even easier for small groups to dominate conversations or create a majority illusion, leveraging ever more realistic bots or agentic AI, raising urgent questions about how to preserve the integrity of the public square. Figuring out who is or is not an authentic user is critical when considering deliberative technology such as collective dialogue systems, bridging systems, and community-driven moderation – not least because of the potentially negative implications for user trust in the technologies, as well as each other.

Proof of humanity (PoH) systems, sometimes called proof-of-personhood, are mechanisms to distinguish between human and automated entities in digital spaces. Sometimes PoH systems operate by connecting an online account to offline identity credentials, such as government issued identity cards or passports. In other cases, no personal information needs to be disclosed directly. "Personhood credentials" (PHCs), are digital credentials that enable users to validate that they are humans to online services, issued by institutions such as governments or other trusted intermediaries. Personhood credentials can be stored digitally on individuals' devices and verified through zero-knowledge proofs, discussed further below (Adler et al. 2024).

The pursuit of reliable methods to verify human presence is fraught with tradeoffs, pitting the imperative of safeguarding against manipulation against principles of privacy, anonymity, and individual autonomy. To complicate matters further, bots (automated agents) and humans are not a simple binary, but exist on a spectrum. For example, automated agents are not always entirely automated; we observe hybrid accounts in which individuals have agents post for them at certain times while operating accounts directly at others (Woolley 2023). Future technologies will likely further blur this distinction between humans and bots, such as "Generative Ghosts," digital avatars designed to live on after a person has passed away (Morris and Brubaker 2024).

## 4.1 Challenges with existing Proof of Humanity Systems

There are numerous challenges to developing proof of humanity systems. For example, virtually any indicator that one might choose is unlikely to be universal – for example, some biometric systems rely on scans of human irises, but not everyone has an iris. Others relate to adoption: in the case of validating a specific identity, someone has to devise, implement, and validate the methodology for attestation. That is likely to be a government, a large organization, or a standards body. There are tradeoffs for each; there are also groups opposed to or distrustful of each.

The following are risks to consider when implementing proof of humanity systems:

- **Coercion**: Coercion and a lack of viable opt-out options when authenticating personhood negate any claim of valid consent. For example, coercive collection of biometric data can be considered an attack on human dignity and bodily integrity.
- **Discrimination**: Risks of discrimination include both perpetuating existing biases and introducing new forms of prejudice. This can exacerbate societal inequities and make marginalized individuals even more vulnerable.
- **Privacy**: Reduced anonymity facilitates harassment and surveillance, potentially subjecting individuals to targeted monitoring and attacks. Privacy-invasive methods also have more dire consequences in case of data breaches.
- **Security breaches**: Security breaches range from data leaks to identity theft (or sale), and are compounded by the persistent threat of cyberattacks on aging infrastructure.
- **Function creep**: Initial innocuous use can be repurposed – intentionally or unintentionally – for something more nefarious.
- **Errors**: Whether technical or human, malicious or accidental, errors can wreak havoc on proof-of-humanity systems. These vulnerabilities are especially problematic when they lead to disenfranchisement from essential services. In cases where biometric identifiers are used, the immutability inherent in biometric data compounds these risks, leaving individuals vulnerable to permanent exclusion or misidentification.

## 4.2 Risks to Proof of Humanity posed by LLMs

The weaknesses described above may be amplified and augmented by the introduction of LLMs. Several types of malign activity will be made easier through LLMs, including both attacks on model systems and exploitation of models (Blazina 2020; Marchal et al. 2024):

- **Data leakage**: Accounts deliberately use adversarial prompting techniques to extract sensitive information.
- **Discerning the source**: The increasing sophistication of AI-generated content blurs the lines between human and machine-created works. The volume of this content that is easily generated not only has a normalizing effect but makes it increasingly difficult for fact checkers and other systems of accountability to label or verify the source of the content. This inability to discern human sources can lead to a devaluation of real human expression, creativity, and journalistic content.
- **Opinion manipulation**: Accounts try to shape public opinion, create confusion, or distract the public from particular events. These campaigns are typically motivated by financial and/or geopolitical ends.

- **Scam**: Accounts pretend to be individuals that they are not, often for the purposes of forming relationships and then trying to manipulate targets into taking financially harmful actions.
- **Spam**: Accounts try to drive targets to visit or engage with low-quality content or domains. In these cases, the LLMs enable both the creation of the longform articles in the content farm domains, as well as the shorter social media posts that attempt to capture audience attention.

LLMs transform this malicious activity by reducing the costs to operate accounts, increasing the volume of output, raising the quality, persuasiveness, and/or precision of output, and reducing detectability. Diffusion models (that generate images) are also impactful here: it is now possible to create images of the same figure at different points in time, and in different locations, clothing, etc. Plausible persona creation is advancing rapidly. Mechanisms for identifying whether an account is what they claim to be, such as examining past behavior, posts, or broader/prior online presence for signs of manipulation, are going to be rendered increasingly ineffective. For years, reCAPTCHA has been one the main defenses used online against automated bots by certifying proof-of-humanity through interactive puzzles. As LLMs become more sophisticated, particularly at solving image-based puzzles, a study by Plesner et al. concludes that "we are now officially in the age beyond captchas" finding no observable difference between humans and LLMs on captcha-solving challenges (Plesner, Vontobel, and Wattenhofer 2024).

### 4.3 Principled deployment of Proof of Humanity Systems

Proof of humanity systems must consider the risks above to ensure equitable and secure access to the services behind the systems, as well as cultivate user trust in the technology. The following is a list of considerations for principled deployment of PoH systems:

- **Accountability**: A robust legal and regulatory framework is imperative to establish accountability and provide avenues for remedy and redress in case of misuse, error, or data breach.
- **Efficacy**: Efficacy assessments are vital to evaluate the effectiveness of attestation systems. Are they accurate and effective? Against what potential threats? How easy is it to game the system?
- **Inclusivity**: This entails removing barriers to access and use, rigorous testing, engagement with varied communities, and prioritizing fairness of experience.
- **Interoperability**: locking in users to a single platform via the authentication system limits the user's choice and can concentrate power in one platform or ecosystem. Proof of humanity or personhood credentials should be portable, requiring identity attestation systems to be interoperable between credentialing services.
- **Privacy**: Privacy considerations are paramount, demanding strict adherence to principles such as data confidentiality and unlinkability across services to safeguard individuals' personal information.
- **Proportionality**: Public spaces have a wide range of uses, from a place where a community gathers regularly to space anonymous individuals merely transit through. Truly public spaces must accommodate this range of uses. Proportionality dictates the judicious use of data, emphasizing data minimization to ascertain only necessary proofs are requested, and purpose limitation restricts data usage to specific, authorized purposes.
- **Security**: Security measures must encompass fraud and hacking resistance, ensuring sustainability over time to maintain the integrity of digital systems.
- **Self-sovereignty**: Self-sovereign identity is a principled approach to control over one's digital identity, moving away from centralized data ownership to decentralized individual custodianship of data. In the context of PoH, self-sovereignty means consent in exchanging data with a bot, as well as control over what data will be revealed, to whom, how it will be processed, and stored.

There is no one-size-fits-all solution to PoH that mitigates all risks while maintaining a high level of effectiveness. Therefore, proper scoping is essential to balancing risks and benefits for each particular use case, balancing efficacy against the severity of potential harm based on what lies behind the humanity attestation. Relevant questions to ask when considering which method to adopt include: What is the tradeoff between false positives (mistakenly blocking real humans) and false negatives (letting bots in)? What is the threat model? What are the characteristics of the population in question? How interoperable is this PoH system beyond one platform or ecosystem?

Protection from fraud and manipulation is not something that can be achieved by merely verifying humanness. This is in part because no PoH method is 100% foolproof. Additionally, malicious activity can be conducted by humans as well as bots, or by both in concert. This underscores the need to complement PoH systems with other approaches to fraud prevention, such as inspecting patterns of user activity over time, observing networked user behavior, and checking posts for known malicious URLs.

**4.4 Future Research on Proof of Humanity**

Advances in AI necessitate simultaneous advances in digital security techniques such as PoH systems, as well as parallel legal and social challenges associated with adopting these systems. These PoH systems must balance accessibility considerations to avoid being overly stringent such that they exclude people or violate privacy rights. Future research should invest in alternative methods of human verification or machine verification that adapt with the trajectory of LLMs and consider the principles of ethical PoH deployment described above.

One active area of research is developing a robust method for allowing users to reveal information about themselves online without revealing details about who they are, such as verification with anonymity. For example, ongoing work on zero-knowledge proofs such as zk-SNARKs ("What Are Zk-SNARKs?" 2021).

Personhood credentials are one promising application of zero-knowledge proofs for PoH. PHCs aim to strike a balance between maintaining the Internet's commitment to privacy and unrestricted access while providing a means to combat the potential overwhelming increase in AI-powered deceptive activities. Further research can help ensure that PHCs carefully navigate complex trade-offs in ecosystem design, including minimizing data storage and processing, preventing credential transfer or theft, and avoiding the concentration of power in a single issuer, all of which can potentially compromise the system's integrity or user rights if not managed properly. To offer one example from Adler et al., research can explore digital accessibility and literacy to facilitate PHC adoption, including studying effects of pre-installing digital credential management applications on smartphones, exploring the use of password managers for securing PHCs, and investigating ways to normalize PHC use through integration with digital civic activities (Adler et al. 2024).

In order to move toward a global model where identity assurance is required for trusted communication, investment is needed in universal identity attestation standards, identity portability and adaptive PoH systems that can evolve with rapidly-advancing AI.

## 5. Conclusion

Discussions of important political issues online often leave internet users feeling frustrated and disengaged. Online discourse frequently devolves into divisive arguments, lacks diverse perspectives, and fails to foster meaningful debate, at times enabling extremism and dehumanizing speech. Discussing important political issues is rarely a positive experience for most internet users; a poll by Pew found that 55% of people were "worn out" by political discussions online (Blazina 2020). Online discourse is often divisive, unrepresentative, and falls short of meaningful deliberation or dialogue. In parallel, generative

AI poses further disruptive impacts - epistemic, material, and foundational - to our democratic processes and public squares (Summerfield et al. 2024).

Alternative futures are possible. Collective Dialogue Systems provide an example of how productive and meaningful discussions could occur in both formal assemblies and informal online fora. Large online platforms could adopt aspects of these CDS, including bridging algorithms which could reduce the incentives for misleading and divisive content that currently exist. Innovative design in discussion forums could target empowering community members to weigh in on cases requiring moderation. Improvements in proof of humanity systems can further mitigate the ability of inauthentic actors to manipulate discourse.

Productive debate also requires all parties to be good faith actors. Some discussions involve actors who attempt to inflame tensions, scapegoat others, engage in intimidation, or amplify known misinformation. This is especially true when discussing issues germane to historically marginalized groups, who may not wish to engage in communities where such tactics undermine their humanity and dignity. Part of cultivating healthy public squares is de-escalating such conflicts and removing bad faith actors.

## 6. Recommendations for Future Work

Achieving this future is not inevitable, however. It requires active investment and coordination among diverse stakeholders. Chief among these are interdisciplinary research, community-driven moderation, and open source tooling to help grow a diverse community of researchers, technologists, and policy makers seeking to make the digital public square more accessible and useful. In the longer term, regulation for the responsible and ethical use of AI in the public square, and platforms for collaborations between different organizations in society, will ensure that AI and the public square is developed in a way that brings out the best of free societies.

Seizing this opportunity will also demand sustained attention to rapidly-evolving dynamics of user trust in and social acceptance in AI systems, the uptake of which is already complicated by competing cultural narratives about what AI is and the impact it is likely to have, shifting mental models, and varying public levels of technical literacy (Xu et al. 2024). Understanding how, why, when, and how far people trust AI systems is especially critical with deliberative technology, because these opportunities involve bringing AI systems into spaces that are traditionally exclusively human, such as interpersonal conversation or dialogue, on questions with consequential social or political stakes.

These points highlight what kinds of new work might be most fruitful. However, in addition to the *what*, it will also be crucial to consider the *how* and the *who*. The "how" question concerns the choice of methods and principles to be applied in such new work. The "who" question is about matching the relevant tasks to the people most appropriately positioned to perform them.

In considering the "how" question, thoughtfully designed interventions should be paired with careful and systematic evaluation, using an experimental approach to understand and learn from the impacts of new techniques. Doing this rigorously in real-world settings is undeniably challenging. However, the use of standard methods such as field tests, A/B comparisons and paired control groups should go some distance toward providing the kind of information needed.

Transparency is another methodological priority. The introduction of technical innovations should be paired with communication to users, the research community, and the wider public. Scoping such communication may be challenging in commercial settings, given the legitimate need to guard valuable intellectual property. However, some threshold of information is required to assure that public-square participants understand the conditions under which they are engaging, and to allow third parties to assess how innovations are impacting social interaction.

Who is in the most appropriate position to undertake the new work we've outlined? Given the central role that tech companies play in creating and maintaining the platforms that undergird much of the digital public square, some of the relevant work will naturally fall to them and will be part and parcel with making their products more useful. Industry is also well positioned to contribute given its unrivaled access to engineering talent and ability to operate at scale and at pace. However, the role of industry will ideally be complemented by innovations coming from civil society, as currently represented by a range of non-profit groups focusing on deliberative tech and social media design and governance. Collaborations between industry and civil society groups offer an appealing way to balance between the strengths of these two kinds of organizations. Academia stands in a complementary position, well situated to provide evaluations of new technologies, as well as to investigate new concepts and approaches which industry might not otherwise discover.

Another side of the field falls naturally to the public sector. There has recently been increasing legislative attention to social networks and other digital platforms, alongside an emerging regulatory environment for AI. Policymakers should carefully evaluate the role that regulation might play at the interface between AI and the digital public square. Here, as in the adjacent cases, it will be crucial to balance (1) the goals of setting clear standards and establishing a level playing field, and (2) the goals of fostering innovation and maximizing the openness and democratic vibrancy of the digital public square.

A final piece of the picture is the general public. A defining characteristic of the public square is that it is shaped by those who participate. Participant-centered innovation and design is key to all four types of technology discussed. Centering the "public" in the digital public square should be a paramount standard for all future work and investments.

| Near-term | Mid-term | Long-term |
|---|---|---|
| **Open-source tools and platforms for collective dialogue and deliberation.** These tools should be accessible, user-friendly, and adaptable to different contexts and communities. They should be able to incorporate proof-of-humanity technologies in the future while still ensuring privacy and anonymity when desired.<br><br>**Community-driven moderation efforts.** This includes developing AI-powered tools that can assist moderators in identifying and addressing harmful content, summarizing large sets of opinions, routing opinions, and fostering a culture of respect, inclusivity and productive discussion in online communities.<br><br>**Interdisciplinary research to understand social and political factors that influence the success of deliberative technologies.** This research should look at the impact of these technologies on civic engagement, polarization, and social cohesion. It should also explore the ethical implications of using AI in these contexts, issues of fairness and transparency, as well as sociotechnical dynamics of trust and acceptance. | **Explore the use of AI in generating educational content for deliberation and facilitating deliberative conversations**. This could include developing AI-powered tools that can summarize complex information, generate discussion prompts, and facilitate dialogue between participants with different viewpoints.<br><br>**Different approaches to bridging political divides and promoting empathy in online spaces.** This work could develop algorithms that expose users to diverse perspectives, facilitating bridging dialogue, and creating spaces for meaningful interaction between different groups.<br><br>**Proof-of-humanity systems to ensure authentic human participation in online discussions.** This research should focus on incorporating existing proof-of-humanity systems like Worldcoin into deliberative platforms. Open problems include ensuring methods are privacy-preserving and universally accessible. | **Develop policies that regulate responsible and ethical use of AI in the digital public square.** This includes addressing issues of bias and transparency, as well as proof-of-personhood and ensuring that users get notified when AI is used.<br><br>**Create platforms for collaboration between researchers, engineers, policy-makers, and civil society organizations to ensure that the development and deployment of deliberative technologies are aligned with democratic values and serve the public good.** This collaboration should involve diverse stakeholders from different backgrounds and perspectives to ensure that these technologies are inclusive and equitable.<br><br>**Explore the development of AI-enhanced deliberative platforms that can seamlessly integrate multiple languages and cultural contexts.** This could facilitate truly global conversations on complex issues while preserving nuance, context, privacy and anonymity. |

*Figure 5: Specific recommendations for further research.*

## Acknowledgements

We are grateful to Beth Noveck, Dane Gambrell, Yasmin Green, Scott Carpenter, Shira McNamara, William Isaac, Malika Mehrotra, and several other experts and colleagues for their reviews and support on this paper. Thanks to Erick Fletes for his contributions of original graphics designed for this paper. We would also like to thank Maya Agarwal, Spencer Baim, Peter Wiegand, Allen Gunn and all of those who planned and facilitated the convening in April 2024 on AI and the Public Square. The rich insights generated at that event provided the foundations for this paper and we are grateful for all the participants from academia, civil society, technology companies, and government who contributed.